\documentclass[aps,pra,superscriptaddress,twocolumn]{revtex4-2}

\usepackage[utf8]{inputenc}
\usepackage{amsmath, mleftright}
\usepackage{braket}
\usepackage{amsfonts}
\usepackage{amssymb}
\usepackage{lipsum}
\usepackage{dsfont}
\usepackage{stmaryrd}
\usepackage{graphicx}
\usepackage{booktabs}
\usepackage{siunitx}
\usepackage[version=4]{mhchem}
\usepackage[hidelinks]{hyperref}
\usepackage[capitalise]{cleveref}
\usepackage{siunitx}
\usepackage{mathtools}
\usepackage{xcolor}
\usepackage{microtype}
\usepackage{algorithm2e}
\usepackage{algpseudocode}
\usepackage{acro}
\usepackage[normalem]{ulem}

\newcommand{\exc}[3]{\tau_{#2^#1}^{#3^#1}}
\newcommand{\dexc}[6]{\tau_{#2^#1#5^#4}^{#3^#1#6^#4}}
\newcommand{\texc}[9]{\tau_{#2^#1#5^#4#8^#7}^{#3^#1#6^#4#9^#7}}
\newcommand{\antiexc}[3]{\kappa_{{#2}^{#1}}^{{#3}^{#1}}}
\newcommand{\dantiexc}[6]{\kappa_{{#2}^{#1}{#5}^{#4}}^{{#3}^{#1}{#6}^{#4}}}
\newcommand{\tantiexc}[9]{\kappa_{{#2}^{#1}{#5}^{#4}{#8}^{#7}}^{{#3}^{#1}{#6}^{#4}{#9}^{#7}}}

\newcommand{\vscf}{\ket{\Phi_{\boldsymbol{i}}}}

\newcommand{\crea}[2]{a^{#1\mspace{-0.5mu}\raisebox{0.3ex}{$\scriptstyle\dagger$}}_{\mspace{-0mu}#2^{#1}}}

\newcommand{\anni}[2]{a^{#1}_{\mspace{-0mu}#2^{#1}}}

\newcommand{\approptoinn}[2]{\mathrel{\vcenter{
    \offinterlineskip\halign{\hfil$##$\cr
    #1\propto\cr\noalign{\kern2pt}#1\sim\cr\noalign{\kern-2pt}}}}}

\newcommand{\stkout}[1]{\ifmmode\text{\sout{\ensuremath{#1}}}\else\sout{#1}\fi}

\def\conflictsname{Conflicts of interest}

\newcommand{\auchem}{Department of Chemistry, Aarhus University, DK-8000 Aarhus C, Denmark}
\newcommand{\auphys}{Department of Physics and Astronomy, Aarhus University, DK-8000 Aarhus C, Denmark}
\newcommand{\kvantify}{Kvantify Aps, DK-2300 Copenhagen S, Denmark}

\DeclareAcronym{hf}{
    short = HF ,
    long = Hartree-Fock ,
}
\DeclareAcronym{fci}{
    short = FCI ,
    long = full configuration interaction ,
}
\DeclareAcronym{vqe}{
    short = VQE ,
    long = variational quantum eigensolver ,
}
\DeclareAcronym{vadapt}{
    short =  vADAPT-VQE ,
    long = vibrational ADAPT-VQE ,
}
\DeclareAcronym{algo}{
    short =  FAST-VQE,
    long = Fermionic Adaptive Sampling Theory VQE ,
}
\DeclareAcronym{en}{
    short = EN ,
    long = Epstein-Nesbet ,
}
\DeclareAcronym{enlm}{
    short = HSCI ,
    long = Heuristic Selected CI,
}
\DeclareAcronym{dsgn}{
    short = HG ,
    long = Heuristic Gradient ,
}
\DeclareAcronym{nisq}{
    short = NISQ ,
    long = Noisy Intermediate-Scale Quantum  ,
}
\DeclareAcronym{sci}{
    short = SCI ,
    long = Selected Configuration Interaction,
}
\DeclareAcronym{jw}{
    short = JW ,
    long = Jordan-Wigner ,
}
\DeclareAcronym{qeb_adapt_vqe}{
    short = QEB-ADAPT-VQE ,
    long = Qubit Excitation Based ADAPT-VQE ,
}
\DeclareAcronym{uvcc}{
    short = UVCC ,
    long = Unitary Vibrational Coupled Cluster ,
}
\DeclareAcronym{duvcc}{
    short = dUVCC ,
    long = disentangled Unitary Vibrational Coupled Cluster ,
}

\DeclareAcronym{vscf}{
    short = VSCF,
    long = Vibrational Self-Consistent Field
}
\DeclareAcronym{pes}{
    short = PES,
    long = Potential Energy Surface
}
\DeclareAcronym{fvci}{
short = FVCI,
long = Full Vibrational Configuration Interaction
}
\DeclareAcronym{vcisd}{
short = VCISD,
long = Vibrational Configuration Interaction SD
}
\DeclareAcronym{vcisdt}{
short = VCISDT,
long = Vibrational Configuration Interaction SDT
}
\DeclareAcronym{ucc}{
short = UCC,
long = Unitary Coupled Cluster
}
\DeclareAcronym{vcc}{
short = VCC,
long = Vibrational Coupled Cluster
}
\DeclareAcronym{ducc}{
short = dUCC,
long = disentangled Unitary Coupled Cluster
}
\DeclareAcronym{ipr}{
short = IPR,
long = infinite product representation
}

\usepackage[normalem]{ulem}

\begin{document}
    \title{Vibrational ADAPT-VQE: Critical points leads to problematic convergence}

    \author{Marco Majland}
    \thanks{Corresponding author: mm@kvantify.dk}
    \affiliation{\kvantify}
    \affiliation{\auphys}
    \affiliation{\auchem}

    \author{Patrick Ettenhuber}
    \affiliation{\kvantify}

    \author{Nikolaj Thomas Zinner}
    \affiliation{\kvantify}
    \affiliation{\auphys}

    \author{Ove Christiansen}
    \affiliation{\kvantify}
    \affiliation{\auchem}

    \begin{abstract}
        Quantum chemistry is one of the most promising applications for which quantum computing is expected to have significant impact. Despite considerable research in the field of electronic structure, calculating the vibrational properties of molecules on quantum computers remain a relatively unexplored field. In this work, we develop a \ac{vadapt} formalism based on an \ac{ipr} of anti-Hermitian excitation operators of the \ac{fvci} wavefunction which allows for preparing eigenstates of vibrational Hamiltonians on quantum computers.
        In order to establish the \ac{vadapt} algorithm using the \ac{ipr}, we study the exactness of \ac{duvcc} theory and show that \ac{duvcc} can formally represent the \ac{fvci} wavefunction in an infinite expansion.
        To investigate the performance of the \ac{vadapt} algorithm, we numerically study whether the \ac{vadapt} algorithm generates a sequence of operators which may represent the \ac{fvci} wavefunction. Our numerical results indicate frequent appearance of critical points in the wavefunction preparation using \ac{vadapt}. These results imply that one may encounter diminishing usefulness when preparing vibrational wavefunctions on quantum computers using \ac{vadapt} and that additional studies are required to find methods that can circumvent this behavior.
    \end{abstract}

    \maketitle

    \section{Introduction}\label{sec:introduction}
    The simulation of many-body quantum systems, such as molecular systems, on quantum computers, holds promise for achieving great computational advantages~\cite{daley_practical_2022,lee_is_2022,elfving_how_2020, aspuru-guzik_simulated_2005}.
    However, the presence of noise and errors significantly diminishes the available computational resources on quantum devices. For \ac{nisq} devices, hybrid quantum-classical algorithms, like the \ac{vqe},  have demonstrated the potential of quantum computers early on~\cite{anand_quantum_2022,mcclean_theory_2016}. The estimation of properties related to nuclear motion in molecules using quantum computing has garnered attention in recent years~\cite{mcardle_digital_2019,ollitrault_hardware_2020,sawaya_near-_2021,sawaya_resource-efficient_2020, majland_optimizing_2023}. A natural first focus is anharmonic molecular vibrations forming the basic foundation for dynamics and free energies. However, the vibrational structure of molecules differs from the electronic structure and could give rise to both new opportunities as well as new challenges.
    
    An essential aspect of estimating vibrational properties of molecules on a quantum computer is the parametrization of the vibrational ground state. Recent studies using the \ac{vqe} have utilized the unitary equivalent of variational Coupled Cluster theory in a form similar to that employed for electronic structure investigations  \cite{mcardle_digital_2019,ollitrault_hardware_2020,anand_quantum_2022, romero_strategies_2018,lee_generalized_2019,ryabinkin_qubit_2018,ryabinkin_iterative_2020} This, however, leads to deep quantum circuits that are difficult to execute on current noisy hardware. To attempt to mitigate the deep quantum circuits of the UCC Ansatz, the electronic structure community turned towards adaptive methods for constructing the electronic wavefunctions on a quantum computer. Such methods are referred to as ADAPT-VQE algorithms and rely on gradient-based constructions of the wavefunctions. Such methods, however, are reliant on gradient evaluations which impose difficulties in estimating optimal sequences of operators when the algorithms are subject to noise. However, current studies imply that the ADAPT-VQE algorithms yield more compact wavefunctions compared to UCC wavefunctions in noiseless simulations
    ~\cite{grimsley_adaptive_2019,tang_qubit-adapt-vqe_2021,feniou_overlap-adapt-vqe_2023,majland_fermionic_2023,nykanen_mitigating_2022,yordanov_qubit-excitation-based_2021}.
    
    The use of ADAPT-VQE for electronic structure calculations relies on the fact that an infinite product of exponentials of low rank excitation operators can formally parametrize the exact electronic \ac{fci} wavefunction \cite{grimsley_adaptive_2019,evangelista_exact_2019}.
    Evangelista et al. have conducted a comprehensive study on the exactness and associated conditions of such a \ac{ducc} theory within electronic structure theory~\cite{evangelista_exact_2019}. In addition, numerical investigations suggest that the ordered pools selected by the gradient descent method of ADAPT-VQE may not be optimal and the sequence identified may differ from a finite part of the IPR sequence following from dUCC \cite{burton_exact_2023}. However, it is an open question whether the \ac{ducc} and the \ac{vadapt} Ansatz can exactly parametrize an exact vibrational wave function. 
    In this paper, we develop a \ac{vadapt} algorithm through demonstrating that the \ac{fvci} wavefunction may be constructed using as an \ac{ipr} of low rank anti-Hermitian excitation operators. The \ac{ipr} representation is obtained through studying the exactness of the \ac{duvcc} Ansatz analogous to the methods utilized in Ref.~\cite{evangelista_exact_2019}.
    Our work is solely concerned on with the exact representability of the \ac{fvci} using an \ac{ipr} as a means for formulating the \ac{vadapt} algorithm. While optimal orderings of excitation operators in finite product representation of the \ac{duvcc} Ansatz is a very interesting problem, such investigations are outside the scope of the this work. We refer to Refs. \cite{halder_dual_2022,mondal_development_2023} for a more elaborate discussion on the ordering of operators in \ac{ducc} in the context of electronic wave functions.

    In contrast to the electronic structure case, our numerical studies of \ac{vadapt} reveal the emergence of critical points that may impede the convergence of the algorithm. Such critical points have been discussed using the UCC path integral formalism in Ref.~\cite{evangelista_exact_2019} as a theoretical possibility but they have not been observed in practice for the electronic structure case. Furthermore, we numerically investigate the properties of the emerging critical points by varying the parameters of the calculation. We find that the \ac{vadapt} algorithm appears to converge to distinct subspaces of the vibrational Hamiltonian, depending only on the operator pool but not on the order in which the operators are used, indicating larger critical surfaces rather than isolated critical points. 
    
    \section{Theory} \label{sec:background}
    \subsection{Background} \label{sec:uvcc}

We employ a many-mode second quantization formalism for vibrational structure. In this section, we will introduce the relevant theoretical background along with conventional notation as was used in Ref. \cite{christiansen_vibrational_2004}.\\
  Let $\{m_1,m_2,...\}$ denote distinguishable vibrational modes and let $\{p,q,r,...\}$ denote general one-mode functions (modals) for each mode.  Let $\{i,j,k,...\}$ denote occupied modals and $\{a,b,c,...\}$ denote unoccupied modals. The operators $\crea{m_1}{p}$ and $\anni{m_1}{q}$ create and annihilate occupation in one-mode basis functions (modals) indexed by $p^{m_1}$ and $q^{m_1}$ for a given mode $m_1$.  
   The creation and annihilation operators satisfy commutator relations
   \begin{equation}
       [\anni{m_1}{p},\crea{m_2}{q}]= \delta_{m_1m_2} \delta_{p^{m_1}q^{m_2}}
          \end{equation}
          and
          \begin{equation}
       [\crea{m_1}{p},\crea{m_2}{q}]= [\anni{m_1}{p},\anni{m_2}{q}] = 0
   \end{equation}
    Note that the creation and annihilation operators are not harmonic oscillator ladder operators and the modals are not assumed to be harmonic oscillator functions. Rather they are formally general one-mode basis functions. In practice a set of modals may, as in this work, be obtained from \ac{vscf} theory. The creation and annihilation operators are encoded onto a quantum computer using the direct encoding. The optimal choice of encoding depends on the structure of the operators to be encoded. However, the direct encoding provides a reasonable trade-off between qubit and gate counts \cite{ollitrault_hardware_2020,sawaya_resource-efficient_2020}.

  An anti-Hermitian $N$-body excitation operator is defined as
    \begin{equation}
        \kappa_{i^{m_1}j^{m_2}..}^{a^{m_1}
        b^{m_2}...}=\tau_{i^{m_1}j^{m_2}...}^{a^{m_1}b^{m_2}...} - \textrm{h.c.},
    \end{equation}
    with
    \begin{equation}
    \tau_{i^{m_1}j^{m_2}...}^{a^{m_1}b^{m_2}...}=\crea{m_1}{a}\anni{m_1}{i}\crea{m_2}{b}\anni{m_2}{j}\crea{m_3}{c}\anni{m_3}{k}\dots,
    \end{equation}
    denoting the vibrational excitation operators.
As in conventional \ac{vcc} theory\cite{christiansen_vibrational_2004}   the cluster operators are defined as
    \begin{equation}
        T_{1} = \sum_{m}\sum_{a^{m}}t^{m}_{a^{m}}\tau_{i^{m}}^{a^{m}},   
    \end{equation}
    and
    \begin{equation}
        \begin{split}
        T_{2} &= \sum_{m_1 < m_2} \sum_{a^{m_1} b^{m_2}} t_{a^{m_1} b^{m_2}}^{m_1m_2}\tau^{a^{m_1}b^{m_2}}_{i^{m_1}j^{m_2}},\\
        \end{split}
    \end{equation}
where the cluster amplitudes are real parameters. 
Unlike the conventional VCC ansatz of the wave function as $\ket{\Psi_{\textrm{VCC}}} = e^{T}\vscf$ we will here follow a unitary formulation as presented in Ref. \cite{mcardle_digital_2019} for the vibrational case and in accord with the long history of electronic unitary CC theory.\cite{hoffmann_unitary_1988,bartlett_alternative_1989}
The \ac{uvcc} ansatz accordingly takes the form
    \begin{equation}
        \ket{\Psi_{\textrm{UVCC}}} = e^{T-T^{\dagger}}\vscf,
        \label{eq:uvcc}
    \end{equation}
    where $\boldsymbol{i}$ denotes the indices of the ground state modals, i.e. the modals that are occupied in the reference state. The cluster operator, $e^{T-T^{\dagger}}$, may not be efficiently decomposed into a set of quantum gates. Thus, it is attractive to turn to the so-called disentangled form\cite{evangelista_exact_2019}. 

    \subsection{The infinite product representation of the \ac{fvci} wavefunction}

    The \ac{ipr} is defined as
    \begin{equation}
        \ket{\Psi_\textrm{IPR}}=\prod_{i}^{\infty}e^{t_{\mu_{i}}\kappa_{\mu_{i}}}\vscf
    \end{equation}
    where $\kappa_{\mu_{i}}$ are low rank anti-hermitian excitation operators.
    In order to show that an \ac{ipr} may be used to represent the \ac{fvci} wavefunction, we present a general formulation of the \ac{duvcc} Ansatz in Section \ref{sec:duvcc}. In Section \ref{sec:decomposition}, we show that high-order excitation operators may be decomposed into low rank excitation operators using nested commutator relations. In Section \ref{sec:ipr}, we combine these results to show that the \ac{ipr} may parametrize the \ac{fvci} wavefunction.

    \subsubsection{The disentangled unitary vibrational coupled cluster ansatz}
    \label{sec:duvcc}
        
    The \ac{duvcc} Ansatz is the logical equivalent of the electronic disentangled unitary coupled cluster ansatz\cite{evangelista_exact_2019} and reads
    \begin{equation}
        \ket{\Psi_\textrm{dUVCC}}=\prod_{\mu}e^{t_{\mu}\kappa_{\mu}}\vscf.
        \label{eq:duvcc}
    \end{equation}
    The \ac{duvcc} Ansatz may be interpreted as a first-order Trotter expansion of Eq.~\eqref{eq:uvcc}. However, since Eq.~\eqref{eq:uvcc} does not decompose efficiently into quantum gates, one may consider Eq.~\eqref{eq:duvcc} as an Ansatz in its own right rather than an approximation to an Ansatz.
    Because the excitation and de-excitation operators do not commute, a finite term Trotter expansion of the unitary ansatz would not be exact.

    In order for Eq.~\eqref{eq:duvcc} to be a valuable Ansatz, it must be able to represent any state in the Hilbert space.    
 %\label{sec:exactness_duvcc}  
 %   In order for the \ac{duvcc} Ansatz and the vibrational \ac{adapt} algorithm to converge to the exact wave function, the \ac{duvcc} Ansatz must be able to represent any state in the vibrational Hilbert space. 
 %We will now show that the \ac{duvcc} Ansatz may represent any vibrational state. 
 Let $\ket{\Psi}$ denote a general state,
    \begin{equation}
        \ket{\Psi}=\sum_{\mu}c_{\mu}\ket{\Phi_{\mu}}.
        \label{eq:general_state}
    \end{equation}
    where $\ket{\Phi_{\mu}}$ denotes Hartree products which in second quantization correspond to a product of creation operators, $a^{m_1,\dagger}_{p^{m_1}}a^{m_2,\dagger}_{p^{m_2}} \cdots $ for t. The summation over $\mu$ runs over all possible Hartree-Products. In order for the \ac{duvcc} Ansatz to represent any state in the Hilbert space, the following must be shown,
    \begin{equation}
        U_\textrm{dUVCC}^{-1}\ket{\Psi} = \vscf,
    \end{equation}
    in analogy to the electronic structure proof in Ref.~\cite{evangelista_exact_2019}.
    Thus, the \ac{duvcc} operator must be able to invert the general state to the reference state for a given set of parameters. In this section, we prove this for vibrational wave functions closely following the reasoning of Ref.~\cite{evangelista_exact_2019}.\\
    Consider the general expansion in Eq.~\eqref{eq:general_state}. Each single excited state, 
    \begin{equation}
    \ket{\Phi_{i^{m}}^{a^{m}}}=\crea{m}{a}\anni{m}{i}\vscf 
,
    \end{equation} may be eliminated from $\ket{\Psi}$ by applying the inverse rotation operator $e^{-t_{i^{m}}^{a^{m}}\kappa_{i^{m}}^{a^{m}}}$.
    If $c_{a^{m}}^{{m}}$ is the expansion coefficient of $\ket{\Phi_{i^{m}}^{a^{m}}}$ and $c_{\boldsymbol{i}}$ is the expansion coefficient of $\vscf$ we find
    \begin{align}
        e^{-t_{a^{m}}^{{m}}\kappa_{i^{m}}^{a^{m}}}\ket{\Psi}
        & = (c_{\boldsymbol{i}}\cos(t_{a^{m}}^{{m}}) +c_{a^{m}}^{{m}}\sin(t_{a^{m}}^{{m}})\vscf \nonumber \\
        & + (c_{a^{m}}^{{m}}\cos(t_{a^{m}}^{{m}})-c_{\boldsymbol{i}}\sin(t_{a^{m}}^{{m}}) )\ket{\Phi_{i^{m}}^{a^{m}}}
        +\cdots,
    \end{align}
    where the second term can be eliminated by the choice 
    \begin{equation}
    t_{a^{m}}^{{m}} = \arctan\Big(\frac{c_{a^{m}}^{{m}}}{c_{\boldsymbol{i}}}\Big),
    \end{equation}
     Such a transformation eliminates $\ket{\Phi_{i^{m}}^{a^{m}}}$ from the expansion but may introduce higher order excitations. Importantly, the operation does not introduce any further single excitations to the expansion. Repeatedly, one may perform inverse transformations of each single excitation in the expansion. After such an operation, the expansion no longer contains single excitations but may contain different higher order excitations. By then performing the inverse transformation on the double excitations in the expansion, one may analogously eliminate all such terms. In analogy, these inverse transformations do not introduce any new single or double excitations, but may introduce higher order excitations. By performing the inverse transformations for the entire excitation manifold of the expansion, one arrives at $\vscf$.
    
    \subsubsection{Decomposition of many-body operators}
    \label{sec:decomposition}
    
    The \ac{duvcc} Ansatz contains up to $N$-body operators in order to represent any state in the Hilbert space. However, these operators are very costly to implement. In order to obtain operators which are feasible for \ac{nisq} devices, one may apply rank reduction through recursive commutation relations to decompose the $N$-body operators into one-body and two-body operators \textbf{\cite{halder_dual_2022,mondal_development_2023}}. Thus, using the recursive relations, one may iteratively reduce the high rank operators to low rank operators such that the \ac{duvcc} Ansatz may be written in terms of only low rank operators. As an example, a triples anti-Hermitian excitation operator may be written as
    \begin{equation}
        \kappa_{i^{m_1}j^{m_2}k^{m_3}}^{a^{m_1}b^{m_2}c^{m_3}}=-[\dantiexc{m_1}{i}{a}{m_2}{j}{p},\dantiexc{m_2}{p}{b}{m_3}{k}{c}]
        \label{eq:generalized_operators}
    \end{equation}
    where $p$ denotes an general modal for which $p\notin\{j,b\}$, or in terms of particle-hole exications
    \begin{equation}
        \tantiexc{m_1}{i}{a}{m_2}{j}{b}{m_3}{k}{c}=-[\dantiexc{m_1}{i}{a}{m_2}{j}{ p},[\antiexc{ m_2}{j}{p},\dantiexc{m_2}{j}{b}{m_3}{k}{c}]].
    \label{eq:commutator}
    \end{equation}\\
    Note that these relations may be recursively extended to higher order operators. The proof is presented in the appendix. 

    \subsubsection{Infinite product representation}
    \label{sec:ipr}

    To show that the \ac{ipr} may represent the \ac{fvci} wavefunction, consider the \ac{duvcc} Ansatz 
    \begin{equation}
        \ket{\Psi_\textrm{dUVCC}}=\prod_{\mu}e^{t_{\mu}\kappa_{\mu}}\vscf.
    \end{equation}
    which may represent the \ac{fvci} wavefunction according to Section \ref{sec:duvcc}. Using Eq. \ref{eq:commutator} and the high-order equivalents along with the commutator expansion \cite{evangelista_exact_2019},
    \begin{equation}
        e^{[A,B]}=\lim_{N\rightarrow\infty}\Big(e^{A/\sqrt{N}}e^{B/\sqrt{N}}e^{-A/\sqrt{N}}e^{-B/\sqrt{N}}\Big)^{N},
    \end{equation}
    the \ac{duvcc} Ansatz may be decomposed into an infinite product of low rank excitation operators. Thus, the \ac{ipr} may represent the \ac{fvci} wavefunction. Note that the above decomposition is equivalent to that in Ref. \cite{evangelista_exact_2019}.

    We are only concerned with the existence of an \ac{ipr}. We are not entering into a discussion of the optimal ordering of the operators in a finite \ac{duvcc} Ansatz. While the problem of theoretically investigating orderings of the operators is a very interesting problem, we only use the above result as a prerequisite for establishing the \ac{ipr} of \ac{fvci} wavefunctions and investigate the sequence of operators generated using the \ac{vadapt} algorithm. Thus, we utilize the \ac{vadapt} algorithm to generate the sequence of operators rather than theoretically calculate a sequence of operators for the \ac{duvcc} Ansatz.
    
    \subsection{ADAPT-VQE}
    \label{sec:adapt_vqe}
    The ADAPT-VQE algorithm iteratively builds an Ansatz through successively adding parametrized unitary transformations acting on a reference state $\vscf$. At the $k$'th iteration of the ADAPT-VQE algorithm, the wave function may be written as
    \begin{equation}
        \ket{\Psi^{(k)}} = \prod_{\mu\in\mathcal{A}^{(k)}} e^{t_\mu \kappa_\mu} \vscf, \label{eq:expans}
    \end{equation}
    where $\mathcal{A}^{(k)}$ is the ordered sequence of excitations entering the wave function up to iteration $k$,
    $\kappa_\mu = \tau_\mu - \tau^\dagger_\mu$, with $\tau_\mu$ being an excitation operator, $\mu$
    enumerates the excitation and $\vscf$ denotes the reference state. The operators used to construct the Ansatz is chosen from a pool of operators, $\mathcal{A}=\{\kappa_{\mu}\}$, based on
    an importance metric, $w(\kappa_\mu, \ket{\Psi^{(k)}})$. In standard ADAPT-VQE, the importance metric is the gradient of the energy with respect to the parameter of the operator. The energy of the $k+1$st iteration may be written as
    \begin{equation}
        E^{(k+1)} = \bra{\Psi^{(k)}} e^{ -t_\mu \kappa_\mu} H e^{t_\mu \kappa_\mu} \ket{\Psi^{(k)}},
    \end{equation}
    such that
    \begin{equation}
        \begin{split}
            g_\mu &= \mleft. \frac{\partial E^{(k+1)}}{\partial t_\mu}\mright|_{t_\mu = 0} \\
            &= \langle\Psi^{(k)} | [H,\kappa_\mu] | \Psi^{(k)} \rangle. \label{eq:adaptgrad}
        \end{split}
    \end{equation}
The product over $\mu$ in Eq.(\ref{eq:expans}) is to be interpreted as an ordered set of transformations, which may be i) many and formally infinitely, ii) non-commuting, iii) repeating the same basic operator multiple times.

   \subsection{Path equations}
    In this section, we present a formalism, 
    %the \ac{ucc} path equations, 
    to describe the path of the wave function from a reference state to an arbitrary state. Following~\cite{evangelista_exact_2019}, the path connecting the reference state $\vscf$ and a general state $\ket{\Psi}$ may be described by a path $s^{k}=[\mathcal{A}^{(k)},\{t^{(k)}_\mu\}]$. Thus, the path depends on the operators $\mathcal{A}^{(k)}$ of the Ansatz in iteration $k$ along with the parameters $\{t^{(k)}_\mu\}$ optimized in iteration $k$. Consider the ADAPT-VQE Ansatz which reads
    \begin{equation}
        \ket{\Psi^{(k)}(s^{k})} = U(s^{k})\vscf
    \end{equation}
    where $U(s^{k})=\prod_\mu e^{t^{(k)}_\mu \kappa_\mu}$. The path equation reads
    \begin{align}
        \frac{d}{ds^k}\ket{\Psi^{(k)}(s^{k})} &= \sum_{\mu}\frac{dt^{(k)}_\mu}{ds^k}\frac{dU(s)}{dt^{(k)}_\mu}\vscf\\
        &= \sum_{\mu}\frac{dt^{(k)}_\mu}{ds^k}\ket{\mu}
    \end{align}
    where $\{\ket{\mu}\}$ denotes an orthogonal basis for the Ansatz derivative. In order for $\ket{\Psi^{(k)}(s^{k})}$ to parameterize any state in the Hilbert space, $s^{k}$ must be a solution to the above differential equation. In the case for which $\{\ket{\mu}\}$ is not a complete basis, $s^{k}$ is not a solution to the differential equation and thus the path $s^{k}$ is at a critical point. At this point, one may no longer parametrize any state in the Hilbert space and thus $\ket{\Psi^{(k)}(s^{k})}$ may not converge to the exact solution. The incompleteness of $\{\ket{\mu}\}$ may be investigated by calculating the rank of the Jacobian,
    \begin{equation}\label{eq:jacobian}
        D_{\mu\nu}^{k} = \braket{\Phi_\mu|\nu},
    \end{equation}
    for which the rank must be rank$(D_{\mu\nu}^{k})=\min(|\{\mu\}|,k)$, where $\{\mu\}$ is the set of all possible excitations.

    \section{ Numerical results} \label{sec:compdetails}
    \subsection{Computations}
    
    In the numerical computations, four molecules were investigated, two two-mode molecules, $\textrm{H}_{2}\textrm{O}$ and $\textrm{O}_{3}$ and two four-mode molecules, $\textrm{H}_{2}\textrm{CO}$ and $\textrm{N}_{2}\textrm{CO}$. The data for $\textrm{O}_{3}$ and $\textrm{N}_{2}\textrm{CO} $ are presented in the Supplementary Information. The conclusions drawn in this section for $\textrm{H}_{2}\textrm{CO}$ and $\textrm{H}_{2}\textrm{O}$ are also consistent with $\textrm{O}_{3}$ and $\textrm{N}_{2}\textrm{CO}$. We use the CCSD(F12*)(T)/cc-pVDZ-F12 \ac{pes} constructed
    in a previous work\cite{majland_optimizing_2023} using the adaptive PES construction methods from the MidasCpp program.\cite{MidasCpp2023040}
    %\cite{artiukhin_denis_g_and_christiansen_ove_and_godtliebsen_ian_heide_and_gras_eduard_matito_and_gyh_orffy_werner_and_hansen_mikkel_bo_and_hansen_mads_bottger_and_klinting_emil_lund_and_kongsted_jacob_and_konig_carolin_and_madsen_diana_and_madsen_niels_kristian_and_monrad_kasper_and_schmitz_gunnar_and_seidler_peter_and_sneskov_kristian_and_sparta_manuel_and_thomsen_bo_and_toffoli_daniele_and_zoccante_alberto_and_hojlund_mads_greisen_and_hoyer_nicolai_machholdt_and_jensen_andreas_buchgraitz_midascpp_2022}
    We investigate the dependence of mode couplings for convergence of the energy calculations by using these \acp{pes} with maximum mode coupling level of $n=2$ and $n=3$ below.
    
%    All electronic structure calculations were performed at the CCSD(F12*)(T)/cc-pVDZ-F12 level~\cite{hattigCommunicationsAccurateEfficient2010,petersonSystematicallyConvergentBasis2008} of theory
%    as implemented in the Turbomole~\cite{TurbomoleV7} program suite.
%    The geometry optimization was performed using numerical gradients. The normal mode coordinates were generated using a numerical Hessian using the MidasCpp program \cite{artiukhinMidasCpp2022}, which was also used
%    to construct electronic PESs with the ADGA algorithm. 
The \ac{vscf}, \ac{vcisd}, \ac{vcisdt} and \ac{fvci} computations were performed in MidasCpp.
    The \ac{vscf} calculations were carried out in large B-spline bases where the \ac{vscf} modals were then used as a basis in conventional \ac{fvci} calculations. The ground state \ac{fvci} energy is reasonably well converged in terms of the number of \ac{vscf} modals at four (three) modals for the three-mode (six-mode) systems, and these modest basis set sizes makes our full-space computations possible. \cite{majland_optimizing_2023}
    
    The \ac{vadapt} algorithm calculations were implemented in a quantum computing extension of MidasCpp with interfaces to Qiskit. \cite{treinish_qiskit_2023} This allows for the construction of both the \ac{duvcc} Ansatz along with the \ac{vadapt} Ansatz in integration with the VSCF state and matrix elements from MidasCpp. These constructions were utilized in statevector simulations in Qiskit.\\
    
    In our experiments we are considering different pools of operators, specifically a pool that contains anti-Hermitian singles and doubles excitation operators, and a pool that additionally contained anti-Hermitian triples excitation operators, which we will refer to as the SD pool and the SDT pool, respectively.

    \newcommand{\figsize}{\textwidth}
 
    \subsection{Ground state energies}
 
    \begin{figure*}[ ht]
        \centering
        \includegraphics[width=\figsize]{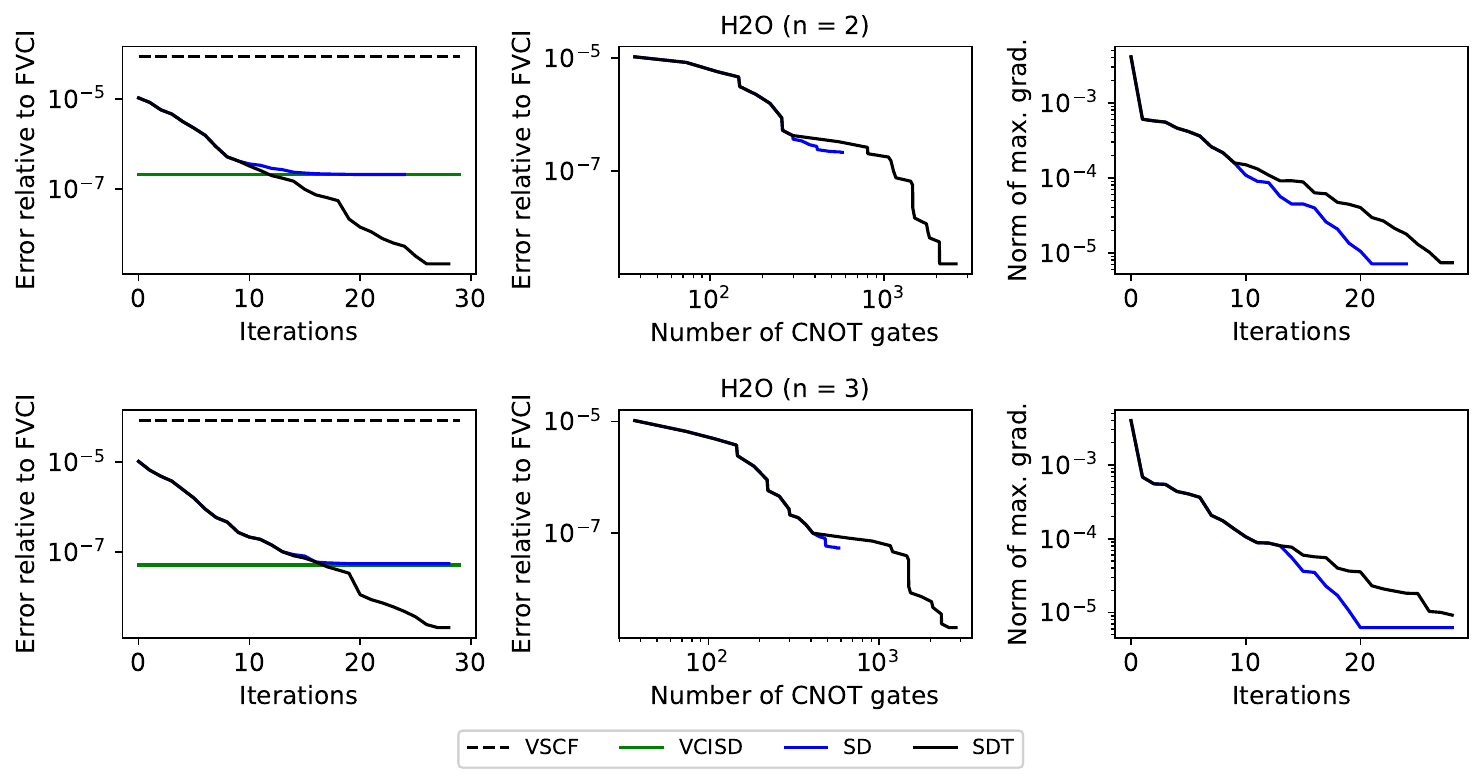}
        \caption{Ground state energy calculations for $\textrm{H}_{2}\textrm{O}$ with mode couplings $n=2$ (upper row) and $n=3$ (lower row). The left figure presents the error relative to the FVCI energy as function of the \ac{vadapt} iterations, $k$. The middle figure presents the error relative to the FVCI energy as a function of the number of CNOT gates in the Ansatz at each iteration $k$ in \ac{vadapt}. The right figure presents the norm of the maximum gradient of the operator pool in \ac{vadapt} as a function of iterations.}
        \label{fig:h2o_2}
    \end{figure*}
  
    \begin{figure*}[ht]
        \centering
        \includegraphics[width=\figsize]{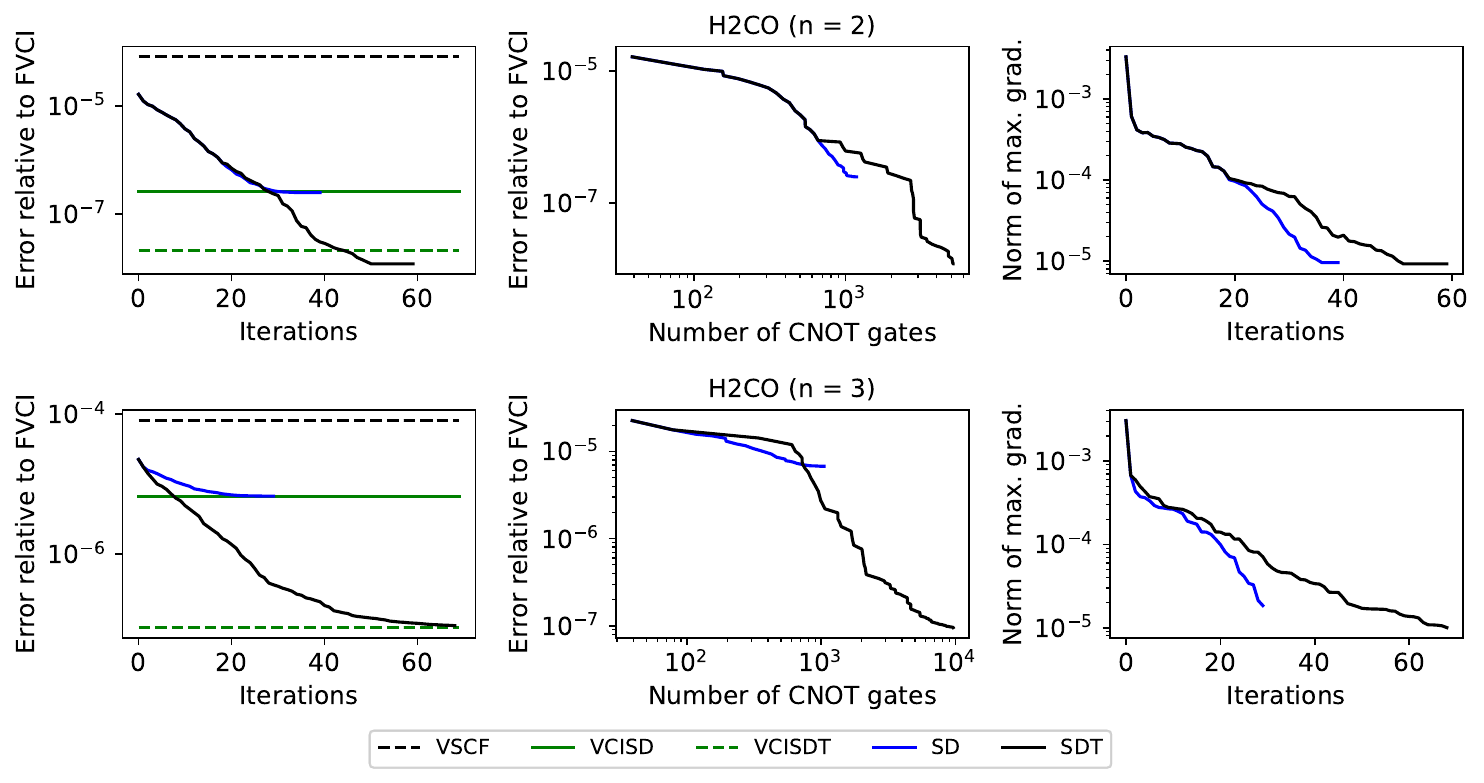}
        \caption{Ground state energy calculations for $\textrm{H}_{2}\textrm{C}\textrm{O}$ with mode couplings $n=2$ (upper row) and $n=3$ (lower row). The left figure presents the error relative to the FVCI energy as function of the \ac{vadapt} iterations, $k$. The middle figure presents the error relative to the FVCI energy as a function of the number of CNOT gates in the Ansatz at each iteration $k$ in \ac{vadapt}. The right figure presents the norm of the maximum gradient of the operator pool in \ac{vadapt} as a function of iterations.}
        \label{fig:h2co_2}
    \end{figure*}

    The error in the ground state energy for $\textrm{H}_{2}\textrm{O}$ with mode-coupling level $n=2$ and $n=3$ is presented in Fig.~\ref{fig:h2o_2}. The left figure demonstrates the convergence of the energy errors for each \ac{vadapt} iteration for both the SD and SDT pools of operators and a horizontal line representing the VCISD energy error. Since both operator pools are theoretically able to represent the FVCI state, it would be expected that both operator pools would be able to converge to the same energy. However, as can be seen in Fig.~\ref{fig:h2o_2}, SD appears to converge to a stationary state with an error relative to the FVCI energy orders of magnitude higher than that of SDT, coinciding roughly with the VCISD energy error. Note that the pattern is independent of $n$.

    The calculations for the ground state energy for $\textrm{H}_{2}\textrm{CO}$ with $n=2$ and $n=3$ are presented in Fig.~\ref{fig:h2co_2}. For $\textrm{H}_{2}\textrm{CO}$ we see the same pattern emerging as we have seen for $\textrm{H}_{2}\textrm{O}$, i.e. \ac{vadapt} calculations using the SD pool seem to converge to the VCISD energy while those with the SDT pool converge around the VCISDT energy, highlighting that this pattern is molecule independent.
    
    \subsection{Critical points}
    
    In order to investigate the convergence behaviour of \ac{vadapt} calculations using the SD and SDT pools, the path equation formalism of Ref.~\cite{evangelista_exact_2019} was utilized. The rank of the Jacobian from Eq.~\eqref{eq:jacobian} was calculated for each iteration, $k$, in the \ac{vadapt} calculations. In the beginning, the rank of the Jacobian increases with each iteration, $\textrm{rank}(D^{k}_{\mu\nu})=k$. However, around the VCISD (VCISDT) energies in the calculations, $\textrm{rank}(D^{k}_{\mu\nu})=k_\textrm{convergence}$ remains constant throughout additional iterations at the same time the norm of the maximum gradient for additional operators tested with the \ac{vadapt} selection criterion is small and becomes constant as shown in Figs.~\ref{fig:h2o_2} and~\ref{fig:h2co_2}. Thus, all \ac{vadapt} calculations using an SD (SDT) pool appear to reach critical points at convergence around the VCISD (VCISDT) energies from which it is not able to escape.
    
    In an attempt to avoid the critical points, several modified algorithms, changing the optimization path, were tested. The first modified algorithm utilizes the operator with maximum gradient (standard \ac{vadapt}) along with a random operator. The second modified algorithm utilizes the two maximum gradient operators. The third modified algorithm utilizes approximate SDT pool containing the standard SD pool along with products of all combinations of one-body and two-body operators, providing decoupled three-body operators, but not including exact three body operators. The fourth modified algorithm uses a pool, denoted SD($k$),  consisting of the standard SD pool with the $k$ most important \emph{exact} three-body operators for SDT, where the importance is determined by the order in which they enter into the wave function when using the full SDT pool.
    
\begin{figure}[ht]
        \centering
        \includegraphics[width=\columnwidth]{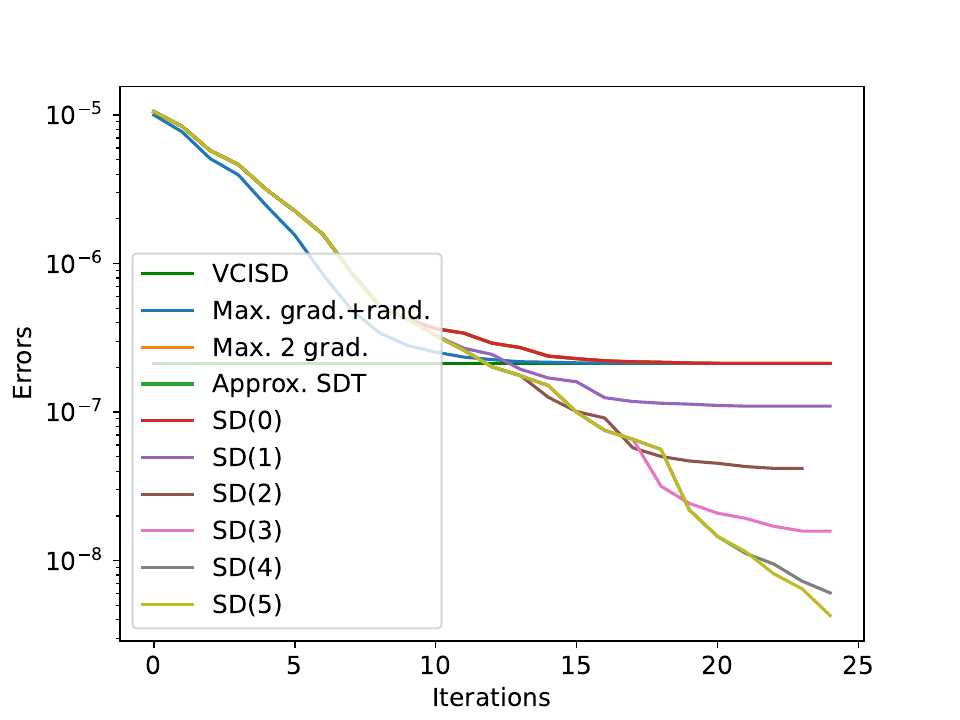}
        \caption{Ground state energy calculations for $\textrm{H}_{2}\textrm{O}$ with $n=2$. The figure presents the error relative to the FVCI energy as function of the \ac{vadapt} iterations, $k$.}
        \label{fig:triples}
    \end{figure}

    In Fig.~\ref{fig:triples}, the calculations for the ground state energy of $\textrm{H}_{2}\textrm{O}$ are shown for each of the modified algorithms. The first, second and third modified algorithms converge to critical points around the VCISD energy. Thus, utilizing operator selections which differ from the "one-operator-gradient-selection" appear to converge to the same critical points. When selecting three-body operators in the operator pool, however, it appears that the critical point close to the VCISD energy is avoided. Nonetheless, the modified SD($k$) algorithm still encounters critical points at different positions along the path, depending on the amount of three-body operators in the respective operator pools.

    In order to test the dependence of the mode couplings on the critical points, a modified Hamiltonian for $\textrm{H}_{2}\textrm{O}$ was investigated. 
    %The modified Hamiltonian amounts to a $n=2$ \ac{pes} for $\textrm{H}_{2}\textrm{O}$. %Let $\boldsymbol{M_{0}}=\{0,1,2\}$ and $\boldsymbol{M_{1}}=\{[0,1],[0,2],[1,2]\}$ denote the mode coupling ranges of the Hamiltonian. 
    A modified n=2 Hamiltonian was introduced as
    \begin{equation}
        H = H^0 + H^1 + H^2 +H^{01}+\alpha(H^{02}+H^{12})%\sum_{m\in\boldsymbol{M_{0}}}h_{m} + \sum_{m,n\in\boldsymbol{M_{1}}}(\delta^{mn}_{01}+\alpha\delta^{mn}_{12}+\alpha\delta^{mn}_{02}) h_{m}h_{n}.
    \end{equation}
    \begin{figure}[ht]
        \centering
        \includegraphics[width=\columnwidth]{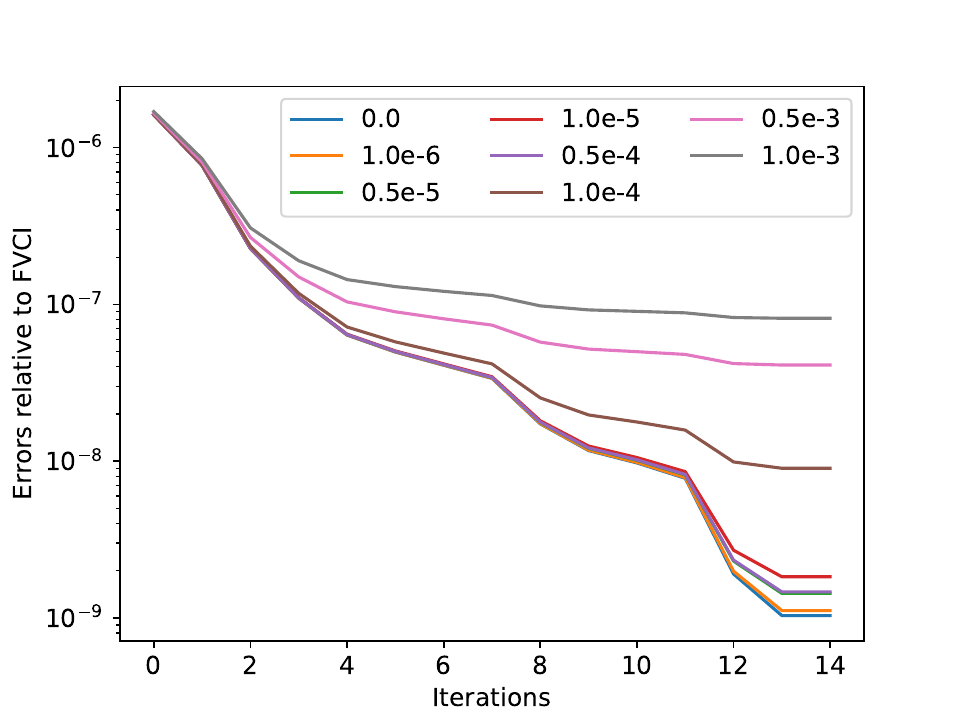}
        \caption{Ground state energy calculations for the $\alpha$-dependent $\textrm{H}_{2}\textrm{O}$ Hamiltonian. The figure presents the error relative to the FVCI energy as function of the \ac{vadapt} iterations, $k$, for different values of the coupling strength, $\alpha$, between modes 0, 1 and 2.}
        \label{fig:mcs}
    \end{figure}
   where $H^m$ includes all terms with only mode $m$ active, and $H^{m_1m_2}$ includes all terms with only modes $m_1,m_2$ active. Since we depart from the $n=2$ Hamiltonian there is no $H^{m_1m_2m_3}$ term. 
   Depending on $\alpha$, one may adjust the coupling between modes 0, 1 and 2. Note that in the case of $\alpha=0$ this means a decoupling of mode $2$: i.e. the exact wave function is a product of a coupled two-mode wave function for modes $0$ and $1$ and an independent mode $2$ factor. 
   
   The results for the modified Hamiltonian are presented in Fig. \ref{fig:mcs}. As would be expected, the precision of the \ac{vadapt} algorithm relative to the \ac{fvci} energy depends on the mode coupling of the \ac{pes}. In the case for which only modes 0 and 1 exhibit a finite coupling strength with mode 2 being completely decoupled, the \ac{vadapt} algorithm converges to the \ac{fvci} energy. With increasing $\alpha$, reflecting increasing coupling strength,  the \ac{vadapt} algorithm convergence fails. 
   This showcase that the standard operator pool leads to critical points in the convergence. 

   We would expect the precise position of the critical points to be somewhat sensitive to noise, but we expect the overall behavior to be the same. Thus, like  different types of operator pools affect the numerical performance of the ADAPT VQE algorithm, with the overall convergence patterns being somewhat similar with exact results not obtained, we expect similar patterns to hold under different levels of noise.
    
    \section{Conclusion} \label{sec:conclusion}
    In this work, we have presented a formalism for the \ac{vadapt} algorithm using the exactness of an \ac{ipr} for the \ac{fvci} wavefunction. These results were obtained based on investigations on the exactness of the \ac{duvcc} Ansatz. Despite the formal exactness of the \ac{ipr} of the \ac{fvci} wavefunction, we numerically demonstrate the appearance of critical points in the convergence of the \ac{vadapt} algorithm in calculating the ground state for several small molecules. Thus, \ac{vadapt} appears to be unable to generate a sequence of operators which converge to the exact vibrational wavefunction.
    
    We investigate different operator pools and mode couplings for different potential energy surfaces in an attempt to escape the critical points. Our findings indicate that the critical points appear to be independent of mode couplings but 
     dependent on the choice of operators pools suggesting critical regions rather than just individual critical points on the accessible manifold. As one includes higher-order excitation operators, the \ac{vadapt} algorithm converges closer to the \ac{fvci} energy. However, in contrast to electronic structure, the \ac{vadapt} algorithm does not converge to the \ac{fvci} energy unless the operator pool is complete. Given this fact,  and the fact that \ac{vadapt} converges to about the same energy error as the corresponding classical VCI theory with the same number of parameters, the potential for computational advantage on the quantum computer using \ac{vadapt} appears out of reach. This conclusion is based on the the fact that the gradient contributions for all operators in the pool needs to be measured to evaluate their importance and the size of the operator pool exhibits the same formal scaling as the scaling of classical CI or CC methods. 
    
     We therefore conclude that further method development is necessary to chart a path towards quantum computational advantage for vibrational structure using near-term quantum devices. 
    
\section*{Supplementary material} \label{sec:supplementary_material}
The supplementary material contains results for $\textrm{O}_{3}$ and $\textrm{N}_{2}\textrm{CO}$.

    \begin{acknowledgments}
    O.C. acknowledges support from the Independent Research Fund Denmark through grant number 1026-00122B. The authors acknowledge funding from the Novo Nordisk Foundation through grant NNF20OC0065479. This work was funded in part by the European Innovation Council through Accelerator grant no. 190124924. %PES calculations were performed at the Centre for Scientific Computing Aarhus (CSCAA).
    \end{acknowledgments}
     
    \section*{Author declarations}
    \subsection*{Conflict of Interest}
    The authors have no conflicts to disclose.
    \subsection*{Author Contributions}
     \textbf{ Marco Majland}
    Conceptualization (equal);
    Data curation (lead);
    Formal analysis (equal);
    Investigation (lead);
    Software (lead);
    Visualization (lead);
    Writing -- original draft (lead);
    Writing -- review \& editing (equal).

    \textbf{Patrick Ettenhuber}
    Conceptualization (equal);
    Formal analysis (equal);
    Supervision;
    Writing -- review \& editing (equal).

    \textbf{Nikolaj Thomas Zinner}
    Conceptualization (equal);
    Formal analysis (equal);
    Funding acquisition (lead);
    Project administration (lead);
    Supervision (lead);
    Writing -- review \& editing (equal).

    \textbf{Ove Christiansen}
    Conceptualization (equal);
    Formal analysis (equal);
    Funding acquisition (lead);
    Project administration (lead);
    Supervision (lead);
    Writing -- review \& editing (equal).
    
    \section*{Data availability}
    The data that support the findings of this study are available from the corresponding author upon reasonable request.
    
    \newpage

    \appendix
    \section{Decomposition of many-body operators}
    To show that an N-body operator may be decomposed into one-body and two-body operators, consider the decomposition of a three-body operator into a nested commutator of one- and two-body operators. For brevity in notation mode indices are denoted $m$,$n$ and $l$ in this appendix. The relation reads
    \begin{equation}
        \kappa_{i^{l}j^{m}k^{n}}^{a^{l}b^{m}c^{n}}=-[\dantiexc{l}{i}{a}{m}{j}{\alpha},\dantiexc{m}{\alpha}{b}{n}{k}{c}]
        \label{eq:generalized_operators2}
    \end{equation}
    where $\alpha$ denotes an unoccupied modal for which $\alpha\notin\{j,b\}$. The relation may be recursively generalized to $N$-body operators. Such a relation also holds for particle-hole operators which reads
    \begin{equation}
        \tantiexc{l}{i}{a}{m}{j}{b}{n}{k}{c}=-[\dantiexc{l}{i}{a}{m}{j}{\alpha},[\antiexc{m}{j}{\alpha},\dantiexc{m}{j}{b}{n}{k}{c}]].
        \label{eq:commutator2}
    \end{equation}\\
    To show the relation for generalized operators, Eq. \ref{eq:generalized_operators2}, consider the commutator
        \begin{align}
        &[\dantiexc{l}{i}{a}{m}{j}{\alpha},\dantiexc{m}{\alpha}{b}{n}{k}{c}] = \\
        &[\dexc{l}{i}{a}{m}{j}{\alpha}-\dexc{l}{a}{i}{m}{\alpha}{j},\dexc{m}{\alpha}{b}{n}{k}{c}-\dexc{m}{b}{\alpha}{n}{c}{k}]\\
        &=[\dexc{l}{i}{a}{m}{j}{\alpha},\dexc{m}{\alpha}{b}{n}{k}{c}]-[\dexc{l}{i}{a}{m}{j}{\alpha},\dexc{m}{b}{\alpha}{n}{c}{k}]\\
        &\quad-[\dexc{l}{a}{i}{m}{\alpha}{j},\dexc{m}{\alpha}{b}{n}{k}{c}]+[\dexc{l}{a}{i}{m}{\alpha}{j},\dexc{m}{b}{\alpha}{n}{c}{k}].
        \end{align}
    Since $l\neq n\neq m$, the above expression reads
    \begin{align}
        &\dexc{l}{i}{a}{n}{k}{c}[\exc{m}{j}{\alpha},\exc{m}{\alpha}{b}]-\dexc{l}{i}{a}{n}{c}{k}[\exc{m}{j}{\alpha},\exc{m}{b}{\alpha}]\\
        &\quad-\dexc{l}{a}{i}{n}{k}{c}[\exc{m}{\alpha}{j},\exc{m}{\alpha}{b}]+\dexc{l}{a}{i}{n}{c}{k}[\exc{m}{\alpha}{j},\exc{m}{b}{\alpha}].
    \end{align}
    Using the identity,
    \begin{equation}
        [\exc{m}{a}{b},\exc{l}{c}{d}] = \delta_{lm}(\exc{l}{c}{b}\delta_{ad}-
        \exc{l}{a}{d}\delta_{bc}),
    \end{equation}
    the expression reads
    \begin{align}
        &\dexc{l}{i}{a}{n}{k}{c}(\exc{m}{\alpha}{\alpha}\overbrace{\delta_{jb}}^{0}-\exc{m}{j}{b}\delta_{\alpha\alpha})
        -\dexc{l}{i}{a}{n}{c}{k}(\exc{m}{b}{\alpha}\overbrace{\delta_{j\alpha}}^{0}-\exc{m}{j}{\alpha}\overbrace{\delta_{b\alpha}}^{0})\\
        &-\dexc{l}{a}{i}{n}{k}{c}(\exc{m}{\alpha}{j}\underbrace{\delta_{b\alpha}}_{0}-\exc{m}{\alpha}{b}\underbrace{\delta_{j\alpha}}_{0})+\dexc{l}{a}{i}{n}{c}{k}(\exc{m}{b}{j}\delta_{\alpha\alpha}-\exc{m}{\alpha}{\alpha}\underbrace{\delta_{jb}}_{0})\\
        &=
        -\texc{l}{i}{a}{m}{j}{b}{n}{k}{c}+\texc{l}{a}{i}{m}{b}{j}{n}{c}{k}\\
        &=-\tantiexc{l}{i}{a}{m}{j}{b}{n}{k}{c}.
    \end{align}
    
    To show the relation for particle-hole operators, Eq. \ref{eq:commutator2}, consider the commutator
    
    \begin{align}        
        &[\antiexc{m}{j}{\alpha}, \dantiexc{m}{j}{b}{n}{k}{c}]=\\
        &[\exc{m}{j}{\alpha}-\exc{m}{\alpha}{j}, \dexc{m}{j}{b}{n}{k}{c}-\dexc{m}{b}{j}{n}{c}{k}]\\
        &=\exc{n}{k}{c}[\exc{m}{j}{\alpha},\exc{m}{j}{b}]
        -\exc{n}{c}{k}[\exc{m}{j}{\alpha},\exc{m}{b}{j}]\\
        &-\exc{n}{k}{c}[\exc{m}{\alpha}{j},\exc{m}{j}{b}]
        +\exc{n}{c}{k}[\exc{m}{\alpha}{j},\exc{m}{b}{j}]\\
        &=\exc{n}{k}{c}(\exc{m}{j}{\alpha}\overbrace{\delta_{jb}}^{0}-\exc{m}{j}{b}\overbrace{\delta_{\alpha j}}^{0})
        -\exc{n}{c}{k}(\exc{m}{b}{\alpha}\delta_{jj}-\exc{m}{j}{j}\overbrace{\delta_{\alpha b})}^{0}\\
        &-\exc{n}{k}{c}(\exc{m}{j}{j}\underbrace{\delta_{\alpha b}}_{0} - \exc{m}{\alpha}{b}\delta_{jj})
        +\exc{n}{c}{k}(\exc{m}{b}{j}\underbrace{\delta_{\alpha j}}_{0} - \exc{m}{\alpha}{j}\underbrace{\delta_{jb})}_{0}\\
        &=-\dexc{m}{b}{\alpha}{n}{c}{k}+\dexc{m}{\alpha}{b}{n}{k}{c}\\
        &=\dantiexc{m}{\alpha}{b}{n}{k}{c}.
    \end{align}
    Using the above expression and Eq. \ref{eq:generalized_operators2}, one obtains Eq. \ref{eq:commutator2}.

\bibliographystyle{jcp}
\bibliography{vibrational_adapt_vqe}

\end{document}